# Axino dark matter from thermal production

**Arnd Brandenburg**[*]

*DESY Theory Group, Notkestrasse 85, D-22603 Hamburg, Germany*
*E-mail:* `Arnd.Brandenburg@desy.de`

**Frank Daniel Steffen**

*DESY Theory Group, Notkestrasse 85, D-22603 Hamburg, Germany*
*E-mail:* `Frank.Daniel.Steffen@desy.de`

ABSTRACT: The axino is a promising candidate for dark matter in the Universe. It is electrically and color neutral, very weakly interacting, and could be—as assumed in this study—the lightest supersymmetric particle, which is stable for unbroken $R$-parity. In supersymmetric extensions of the standard model, in which the strong CP problem is solved via the Peccei-Quinn mechanism, the axino arises naturally as the fermionic superpartner of the axion. We compute the thermal production rate of axinos in supersymmetric QCD. Using hard thermal loop resummation, we obtain a finite result in a gauge-invariant way, which takes into account Debye screening in the hot quark-gluon-squark-gluino plasma. The relic axino abundance from thermal scatterings after inflation is evaluated. We find that thermally produced axinos could provide the dominant part of cold dark matter, for example, for an axino mass of $m_{\tilde{a}} \approx 100\,\text{keV}$ and a reheating temperature of $T_R \approx 10^6\,\text{GeV}$.

---

[*]Heisenberg fellow

## Contents

## 1. Introduction

The strong CP problem and the hierarchy problem motivate symmetries and particles beyond the standard model of particle physics. The strong CP problem can be solved naturally by implementing the Peccei-Quinn (PQ) mechanism [1]. An additional global U(1) symmetry referred to as PQ symmetry broken spontaneously at the PQ scale [2, 3] $f_a \gtrsim 5 \times 10^9$ GeV can explain the smallness (vanishing) of the CP violating $\Theta$-vacuum term in quantum chromodynamics (QCD). The pseudo Nambu-Goldstone boson associated with this spontaneous symmetry breaking is the axion [4], which has not yet been detected. The hierarchy problem is solved in an elegant way by imposing supersymmetry (SUSY) being softly broken at least up to energies accessible in present day accelerator experiments; cf. [5] and references therein. In this way, one can explain the stability of the electroweak (EW) scale against radiative corrections and at the same time why none of the new particles arising from supersymmetric extensions of the standard model has been detected so far. If the standard model supplemented by the PQ mechanism is supersymmetrized [6, 7], in addition to the axion its superpartners—the axino and the saxion—arise, which are not part of the standard spectrum of the minimally supersymmetric standard model (MSSM). Our focus here is on axinos and their cosmological implications.[1]

---

[1] The saxion can also have important cosmological implications [8]. Its decays could, for example, lead to significant entropy production reheating the Universe. However, we assume in this study that the saxion mass is such that saxion effects are negligible.

A key problem in cosmology is the understanding of the nature of cold dark matter in the Universe. With (basically) consistent evidence from dynamics of galaxies and galaxy clusters, gravitational lensing of background galaxies by (super-)clusters of galaxies, insights into the formation and power spectrum of large-scale structures, and the analysis of the anisotropies in the cosmic microwave background (CMB), we believe today that about 22% of the energy density in the Universe resides in cold non-baryonic dark matter [9, 10]. This refers to matter that interacts gravitationally, is electrically and color neutral, and had non-relativistic velocities sufficiently long before the time of matter-radiation equality.

A prominent candidate for cold dark matter is the lightest supersymmetric particle (LSP) provided it is electrically and color neutral and stable on cosmological timescales. The LSP is indeed stable if SUSY is realized with $R$-parity conservation. Depending on the mechanism of SUSY breaking and the form of the superpotential, such an LSP could be, for example, the neutralino, the gravitino, or—and this is the assumption in this work— the axino. The axino is electrically and color neutral and could have a mass such that it moves non-relativistically already long before the time of matter-radiation equality. As its interactions with the MSSM particles are strongly suppressed by the PQ scale $f_a$, the axino is an extremely weakly interacting massive particle. Moreover, as the LSP, the axino destabilizes other $R$-odd SUSY particles. This relaxes constraints in the parameter space of SUSY models [11] and provides a solution [12] to the gravitino problem [13].

Can axinos provide the dominant part of cold dark matter? Assuming the axino is the LSP and stable due to $R$-parity conservation, the following three criteria are crucial to answer this question: (i) Does the relic axino abundance match the observational data on the abundance of cold dark matter? (ii) Is the value of the axino mass large enough to ensure axinos being sufficiently cold (i.e. non-relativistic sufficiently long before matter-radiation equality) so that the formation and power spectrum of large-scale structures can be explained? (iii) Does the abundance of the primordial light elements (D, $^3$He, $^4$He, Li) remain unaffected by the production of axinos? This nucleosynthesis constraint refers to a potential increase of the energy density at the time of primordial nucleosynthesis due to the presence of axinos. Such an increase would accelerate the expansion of the Universe and cause an earlier freeze out of species. As a result, the ratio of protons to neutrons would change and affect the abundance of the light elements. In addition, late decays of the next-to-lightest SUSY particle (NLSP) into axinos are dangerous since accompanying photons or hadronic showers from these NLSP decays could destroy the light elements.

The three criteria above have already been addressed by Covi et al. [14] for axino production in thermal and non-thermal processes. However, the thermal production of axinos has only been estimated in a gauge-dependent and cutoff-dependent framework [14]. In our study we concentrate on the computation of the relic axino abundance from thermal production in the early Universe. In particular, we derive the thermal axino production rate consistently in leading order of the gauge coupling in a proper gauge-invariant treatment within thermal field theory. The axino mass is treated as a free parameter as it depends strongly on the SUSY breaking mechanism and the form of the superpotential [15, 16, 17]. The nucleosynthesis constraints concerning the NLSP decays have already been studied thoroughly for two NLSP candidates, a bino-dominated lightest neutralino [18, 14] and the lighter stau [11]. As the obtained constraints depend very much on the properties of the NLSP—its composition, its mass relative to the one of the axino, and its coupling to the axino—we summarize only the general aspects of these constraints.

The relic axino abundance depends on the cosmic scenario considered. We assume that inflation governed the earliest moments of the Universe; cf. [19, 20, 21] and references therein. Accordingly, any initial population of axinos was diluted away by the exponential expansion during the slow-roll phase. The subsequent reheating phase leads to a Universe of temperature $T_R$. We compute the axino abundance produced in thermal reactions after the reheating phase. We do not consider non-thermal axino production mechanisms possibly present during the reheating phase. Such mechanisms are strongly model dependent as can be seen from studies of gravitino production during inflaton decays [22].

The relative importance of the various axino production mechanisms is governed by the reheating temperature $T_R$ and the mass spectrum of the superparticles. As the PQ symmetry is restored at $f_a$, we consider only values of $T_R$ up to the PQ scale. Another important scale is the temperature $T_D$ at which axinos decouple from the thermal bath. Since the axino interactions are suppressed by the PQ scale $f_a$, the axinos decouple early, i.e. at high temperatures. Rajagopal, Turner, and Wilczek [16] have estimated a decoupling temperature of $T_D \approx 10^9$ GeV, which is consistent with our computations. For $T_R > T_D$, there has been an early phase in which axinos were in thermal equilibrium with the thermal bath. The axino number density is then given by the equilibrium number density of relativistic Majorana fermions while contributions from other processes are negligible. For $T_R < T_D$, the axinos are out of thermal equilibrium so that the production mechanisms have to be considered in detail. Such dedicated investigations have been performed by Covi et al. [14, 23]. In their studies it has been found that axino production in thermal reactions in the hot MSSM plasma governs the relic axino abundance for $T_R \gtrsim 10^4$ GeV. For smaller reheating temperatures, axino production from decays of particles out of equilibrium and, in particular, from NLSP decays is dominant, whose relative importance depends on the masses of the particles involved. Focussing here on axino production in thermal reactions, our results are relevant for $f_a > T_R \gtrsim 10^4$ GeV.

The thermal production of axinos within SUSY QCD has been considered in [14]. Four of the ten production processes are logarithmically singular due to the exchange of massless gluons. In [14] these singularities have been regularized by introducing an effective gluon mass by hand. While this can be considered a reasonable first step, the hard thermal loop (HTL) resummation technique [24] together with the Braaten-Yuan prescription [25] offers a more rigorous tool for dealing with such situations. Using these techniques, we compute the thermal production of axinos within SUSY QCD to leading order in the gauge coupling. The analogous improvement in the computation of thermal gravitino production has lowered the corresponding yield by a factor of three [26].

The observational data on the abundance of cold dark matter in the Universe [9, 10] imply an upper bound on the relic axino abundance. This in turn leads to an upper bound on $T_R$, which depends on the axino mass. High values of $T_R$ seem to be mandatory, for example, for the generation of the baryon asymmetry through thermal leptogenesis [27]. We will investigate if axino dark matter and thermal leptogenesis can coexist.

The paper is organized as follows. In Sec. 2 we review the properties of axions, saxions, and axinos relevant for our investigation. In Sec. 3 we compute the thermal axino production rate in SUSY QCD to leading order in the gauge coupling using the HTL-resummed gluon propagator. The relic abundance of axinos from thermal production in the early Universe is computed in Sec. 4. We discuss axino dark matter scenarios and other cosmological implications of the obtained axino density in Sec. 5. For completeness, the HTL-resummed gluon propagator is provided in the Appendix.

## 2. The axino and its interactions

In this section we briefly review the strong CP problem, the solution proposed by Peccei and Quinn, and the properties of the axion arising in realizations of this solution. The axino and the saxion are obtained by supersymmetrizing axion models. We summarize the axino and saxion properties relevant for our investigation. We discuss the axino interactions and provide the Lagrangian of those most relevant for our investigation. More detailed discussions on axino properties can be found in [16, 14] and references therein.

The isoscalar $\eta'$ meson being too heavy to qualify as a pseudo-Goldstone boson of a spontaneously broken axial $U(1)_A$ is the well-known $U(1)$ problem of the strong interactions. This problem has an elegant solution in QCD based on non-trivial topological properties [28]. The $\eta'$ meson mass can be understood as a consequence of the Adler-Bell-Jackiw anomaly [29] receiving contributions from gauge field configurations with non-zero topological charge such as instantons [30]. The existence of such configurations implies the additional $\Theta$-vacuum term in the QCD Lagrangian [28, 31]

$$\mathcal{L}_\Theta = -\Theta \frac{g^2}{32\pi^2} G^a_{\mu\nu} \tilde{G}^{a\mu\nu}, \tag{2.1}$$

where $g$ is the strong coupling and $G^a_{\mu\nu}$ is the gluon field strength tensor, whose dual is given by $\tilde{G}^a_{\mu\nu} = \epsilon_{\mu\nu\rho\sigma} G^{a\rho\sigma}/2$. For any value of $\Theta \neq n\pi$ with $n \in \mathbb{Z}$, (2.1) violates the discrete symmetries P, T, and CP. Such violations have not been observed in strong interactions and experiments on the electric dipole moment of the neutron give an upper bound of $|\Theta| < 10^{-9}$. Within QCD, $\Theta = 0$ seems natural based on the observed conservation of the discrete symmetries. However, once QCD is embedded in the standard model of strong and electroweak interactions—with CP violation being experimental reality—the value of $\Theta$ and the argument of the determinant of the quark mass matrix $M$—two *a priori* unrelated quantities—must cancel to an accuracy of $10^{-9}$ in order to explain the experimental data:

$$|\bar{\Theta}| \equiv |\Theta - \text{Arg} \det M| < 10^{-9} . \tag{2.2}$$

This is the strong CP problem (cf. also [32] and references therein).

The elegant solution of the strong CP problem suggested by Peccei and Quinn [1] requires to extend the Standard Model with an additional global chiral $U(1)$ symmetry—the PQ symmetry $U(1)_{PQ}$—broken spontaneously at the PQ scale $f_a$. The corresponding pseudo Nambu-Goldstone boson is the axion $a$ [4], which couples to gluons such that the chiral anomaly in the U(1)$_{PQ}$ current is reproduced,

$$\mathcal{L}_{agg} = \frac{a}{f_a/N} \frac{g^2}{32\pi^2} G^a_{\mu\nu} \tilde{G}^{a\mu\nu} \, , \tag{2.3}$$

where the color anomaly $N$ of the PQ symmetry depends on the axion model as discussed below. This interaction term (2.3) together with the vacuum term (2.1) for $\Theta \to \bar{\Theta} = \Theta - \text{Arg}\det M$ provide the axion field with an effective potential $V_{\text{eff}}$ at low energies. This solves the strong CP problem as the coefficient of the CP violating $G\tilde{G}$ term becomes dynamical and vanishes for the value $\langle a \rangle = -\bar{\Theta} f_a/N$ at which $V_{\text{eff}}$ has its minimum.

Because of the chiral U(1)$_{PQ}$ anomaly, the axion receives a mass from QCD instanton effects [33] reflecting the curvature of the effective axion potential $V_{\text{eff}}$ at low energies

$$m_a^2 = \frac{m_u m_d}{(m_u + m_d)^2} \left(\frac{f_\pi m_\pi}{f_a/N}\right)^2 \, , \tag{2.4}$$

where $m_u$ ($m_d$) is the mass of the up (down) quark and $f_\pi$ and $m_\pi$ are respectively the decay constant and mass of the pion. While the original PQ proposal [1] assumed $f_a$ to be of the order of the Fermi scale, particle physics experiments, astrophysical observations, and cosmological arguments provide phenomenological limits that point to a significantly higher value of the PQ scale (cf. [2, 3] and references therein)

$$f_a/N \gtrsim 5 \times 10^9 \, \text{GeV} \, . \tag{2.5}$$

Accordingly, the mass of the hypothetical axion must be very small, $m_a \lesssim 10^{-2}$ eV. In particular, with the axion interactions being strongly suppressed, the axion itself would be sufficiently stable and invisible to provide the cold dark matter in the Universe [34].

The two most popular classes of phenomenologically viable axion models are the hadronic or KSVZ axion models [35] and the DFSZ axion models [36]. In the KSVZ schemes at least one additional heavy quark is introduced which couples directly to the axion while all other fields do not carry PQ charge. Thus, the axion interacts with ordinary matter through the anomaly term from loops of this new heavy quark. Integrating out the heavy quark loops, one obtains the effective dimension-5 coupling of axions to gluons given in (2.3) with $N = 1$. In particular, couplings of the axion to standard model matter fields (such as the couplings of the axion to the light quarks) are suppressed by additional loop factors. In the DFSZ schemes no additional heavy quarks are introduced. Instead, the standard model matter fields and two Higgs doublets carry appropriate PQ charges such that the axion also couples directly to the standard model quark fields. Again, at low energies the axion-gluon-gluon interaction (2.3) arises, but now with $N = 6$. For more details on these axion models, we refer to the reviews [37].

By supersymmetrizing the standard model supplemented by the PQ mechanism, which is possible for both the KSVZ and DSFZ axion models, all fields are promoted to supermultiplets [6, 7]. In particular, the axion is promoted to the axion chiral supermultiplet. Accordingly, in addition to the ordinary MSSM particles and the pseudo-scalar ($R = +1$) axion $a$, the scalar ($R = +1$) saxion $s$ and the spin-1/2 ($R = -1$) axino $\tilde{a}$ appear in the particle spectrum. The axino is the fermionic partner of the axion and the saxion, and thus an electrically and color neutral chiral Majorana fermion. At high energies at which SUSY is unbroken, the masses of the saxion, axion, and axino are degenerate. This changes once SUSY is broken. Being a scalar, the saxion behaves as the other scalar superpartners and acquires a soft-mass term so that the saxion mass is of order of the soft SUSY breaking scale $m_{\text{soft}} = \mathcal{O}(\text{TeV})$. In contrast, the mass of the axino $m_{\tilde{a}}$ depends strongly on the choice of the superpotential and the SUSY-breaking scheme [15, 16, 17]; cf. also [14]. Depending on the model, $m_{\tilde{a}}$ can range between the eV and the GeV scale. As mentioned in the Introduction, we will assume that the axino is the LSP and stable due to $R$-parity conservation. However, we will treat the precise value of the axino mass $m_{\tilde{a}}$ as a free parameter because of its model dependence.

In this work we concentrate on the axino-gluino-gluon interactions obtained from supersymmetrization of (2.3). They are given by the effective dimension-5 interaction term

$$\mathcal{L}_{\tilde{a}\tilde{g}g} = i \frac{g^2}{64\pi^2 (f_a/N)} \bar{\tilde{a}} \gamma_5 [\gamma^\mu, \gamma^\nu] \tilde{g}^a G^a_{\mu\nu} , \qquad (2.6)$$

with the gluino $\tilde{g}$ being the fermionic superpartner of the gluon and $N$ being again the number of quarks with PQ charge. This Lagrangian describes the axino-gluino-gluon 3-vertex and the axino-gluino-gluon-gluon 4-vertex used in our computation of the thermal axino production rate.

For the couplings of axinos to the EW gauge bosons and their superpartners, additional model dependent coefficients $C_{aYY}$ and $C_{aWW}$ appear in front of the corresponding dimension-5 operators. At high energies, at which the leptons can be considered massless, these interactions can be rotated such that $C_{aWW} = 0$ [14]. The remaining interaction between the U(1)$_Y$ field strength and the axion multiplet contains, for example, the axion-photon-photon vertex which gives rise to the Primakov process. However, the corresponding axino interactions with the hypercharge multiplet are subdominant compared to those from (2.6) due to the smaller coefficient $g_Y^2 \, C_{aYY}$ (provided $C_{aYY}$ is not too large) and due to the fact that U(1)$_Y$ is Abelian which implies a smaller number of production channels [14]. In the KSVZ scheme, an electric charge of the heavy quarks of $e_Q = 0, -1/3, +2/3$ gives $C_{aYY} = 0, 2/3, 8/3$ respectively.

In addition to the dimension-5 interactions discussed above, there are (effective) dimension-4 interactions between the axion supermultiplet and the MSSM matter fields, which are more model dependent. In the KSVZ models, such interactions occur only at the two-loop level, i.e. at the one-loop level in the effective theory, and thus are suppressed by loop factors [23, 11]. In the DSFZ models, the axino can mix with the other neutralinos (through the coupling leading to the $\mu$-term) and axino-matter interactions can occur already at the tree level [38]. Independently of the model, the mass dimension of the axino-matter couplings implies that contributions to the axino yield from reactions involving these couplings are suppressed by a factor of $m/T$ at high temperatures, where $m$ is the mass of the corresponding matter field. Indeed, for squark and gluino masses around the TeV scale, the axino-quark-squark interaction within the KSVZ scheme has been found to give irrelevant contributions to the axino yield for $T_R$ above the TeV scale [23].

In summary, concentrating more on the KSVZ models, in which the axino does not mix with the neutralinos,[2] the axino-gluino-gluon interaction (2.6) governs the thermal axino production at high temperatures and the corresponding yield for the range of reheating temperatures, $f_a/N > T_R \gtrsim 10^4$ GeV, considered in this work. Here the other (more model dependent) axino interactions can be neglected as they are subdominant. Nevertheless, these interactions are crucial for the nucleosynthesis constraints since the dominant decay channels of a neutral (charged) NLSP into axinos involve the axino interactions with the hypercharge multiplet (matter fields) [23, 11]; see also Sec. 5.

## 3. Thermal axino production rate

In this section we calculate the thermal production rate of axinos with energies $E \gtrsim T$ at high temperatures from the axino-gluino-gluon interaction (2.6) within SUSY QCD. We use the HTL resummation [24] and the Braaten-Yuan prescription [25] to treat divergencies due to $t$-channel or $u$-channel exchange of soft gluons. This allows us to obtain the production rate to leading order in the strong coupling in a gauge-invariant way without introducing *ad hoc* cutoffs. The exchange of soft fermions does not lead to divergencies so that the use of the HTL resummation is not necessary for soft gluino exchange. For hard axino energies, $E \gtrsim T$, there is also no need to consider HTL-resummed axino-gluino-gluon vertices.

For processes involving $t$-channel ($u$-channel) gluon exchange, we follow the prescription of Braaten and Yuan [25]. We introduce a momentum scale $k_{\text{cut}}$ such that $gT \ll k_{\text{cut}} \ll T$ in the weak coupling limit, $g \ll 1$. This separates soft gluons with momentum transfer of order $gT$ from hard gluons with momentum transfer of order $T$. In the region of soft momentum transfer, $k < k_{\text{cut}}$, the HTL-resummed gluon propagator is used which takes into account Debye screening in the quark-gluon-squark-gluino plasma (QGSGP). The corresponding soft contribution to the axino production rate is obtained from the imaginary part of the axino self-energy with the ultraviolet (UV) momentum cutoff $k_{\text{cut}}$. In the region of hard momentum transfer, $k > k_{\text{cut}}$, the bare gluon propagator can be used. The corresponding hard contribution is then conveniently obtained from the matrix elements of the $2 \to 2$ scattering processes with the infrared (IR) momentum cutoff $k_{\text{cut}}$. While both, the hard and soft contribution, depend logarithmically on $k_{\text{cut}}$, the sum of both is independent of this artificial momentum scale. This sum gives the finite result for the contribution from processes involving $t$-channel ($u$-channel) gluon exchange. The remaining contributions from processes not involving $t$-channel ($u$-channel) gluon exchange can be obtained from naive perturbation theory, which is equivalent to setting $k_{\text{cut}} = 0$.

### 3.1 Hard contribution

The ten $2 \to 2$ scattering processes involving the axino-gluino-gluon interaction (2.6) are those shown in Fig. 3.1. The corresponding squared matrix elements are listed in Table 1. Working in the high-temperature limit, $T \gg m_i$, the masses of all particles involved have been neglected. Sums over initial and final spins have been performed. For quarks and

---

[2]In DFSZ models, the mixing of the axino with the other neutralinos can be non-negligible and affect the thermal production of axinos.

- A: $g^a + g^b \to \tilde{g}^c + \tilde{a}$

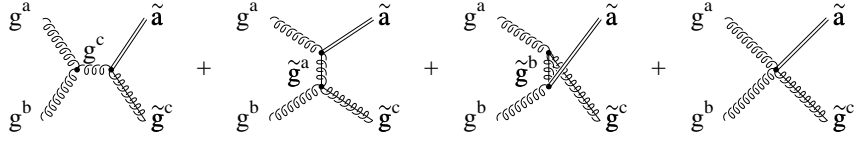

- B: $g^a + \tilde{g}^b \to g^c + \tilde{a}$ (crossing of A)
- C: $\tilde{q}_i + g^a \to \tilde{q}_j + \tilde{a}$ 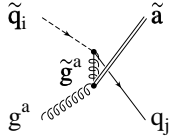
- D: $g^a + q_i \to \tilde{q}_j + \tilde{a}$ (crossing of C)
- E: $\bar{\tilde{q}}_i + q_j \to g^a + \tilde{a}$ (crossing of C)
- F: $\tilde{g}^a + \tilde{g}^b \to \tilde{g}^c + \tilde{a}$ 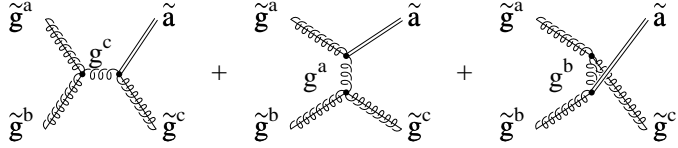
- G: $q_i + \tilde{g}^a \to q_j + \tilde{a}$ 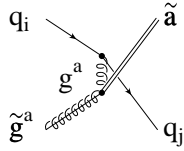
- H: $\tilde{q}_i + \tilde{g}^a \to \tilde{q}_j + \tilde{a}$ 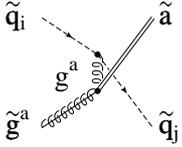
- I: $q_i + \bar{q}_j \to \tilde{g}^a + \tilde{a}$ (crossing of G)
- J: $\tilde{q}_i + \bar{\tilde{q}}_j \to \tilde{g}^a + \tilde{a}$ (crossing of H)

**Figure 1:** The $2 \to 2$ processes for axino production in SUSY QCD.

squarks the contribution of a single chirality is given. The squared matrix elements are expressed in terms of the Mandelstam variables

$$s = (P_1 + P_2)^2 \qquad \text{and} \qquad t = (P_1 - P_3)^2, \tag{3.1}$$

where the particle four-momenta $P_1$, $P_2$, $P_3$, and $P$ refer to the particles in the order in which they are written down in the column "process $i$" of Table 1. To compute the hard contribution to the thermal axino production rate, we divide the ten processes into three classes depending on the number of bosons and fermions in initial and final state involved in addition to the axino: A, C, and J are BBF processes with two bosons in the initial and a fermion in the final state; correspondingly, B, D, E, and H are BFB processes, and F, G, and I are FFF processes. Only the processes B, F, G, and H involve the exchange of a gluon in the $t$-channel ($u$-channel) and therefore contribute to the logarithmic cutoff dependence. Denoting the axino production rate per unit volume by $\Gamma_{\tilde{a}}$, the hard contribution to the axino production rate can be written as follows (cf. [39, 40]),

$$\frac{d\Gamma_{\tilde{a}}}{d^3p}\bigg|_{\text{hard}} = \frac{1}{(2\pi)^3 2E} \int \left[\prod_{i=1}^{3} \frac{d^3 p_i}{(2\pi)^3 2E_i}\right] (2\pi)^4 \delta^4(P_1 + P_2 - P - P_3)$$
$$\times \left(f_{BBF} |M_{BBF}|^2 + f_{BFB} |M_{BFB}|^2 + f_{FFF} |M_{FFF}|^2\right) \Theta(|\mathbf{p}_1 - \mathbf{p}_3| - k_{\text{cut}}) . \quad (3.2)$$

**Table 1:** Squared matrix elements for axino ($\tilde{a}$) production in two-body processes involving quarks ($q_i$), squarks ($\tilde{q}_i$), gluons ($g^a$), and gluinos ($\tilde{g}^a$) in the high-temperature limit, $T \gg m_i$. The results are given for a single chirality of quarks and squarks, for the specified choice of colors, and summed over spins in the initial and final state. $f^{abc}$ and $T^a_{ji}$ are the usual SU(3) color matrices.

| | process $i$ | $|\mathcal{M}_i|^2 / \frac{g^6}{128\pi^4(f_a/N)^2}$ |
|---|---|---|
| A | $g^a + g^b \to \tilde{g}^c + \tilde{a}$ | $4(s + 2t + 2\frac{t^2}{s})|f^{abc}|^2$ |
| B | $g^a + \tilde{g}^b \to g^c + \tilde{a}$ | $-4(t + 2s + 2\frac{s^2}{t})|f^{abc}|^2$ |
| C | $\tilde{q}_i + g^a \to q_j + \tilde{a}$ | $2s|T^a_{ji}|^2$ |
| D | $g^a + q_i \to \tilde{q}_j + \tilde{a}$ | $-2t|T^a_{ji}|^2$ |
| E | $\bar{\tilde{q}}_i + q_j \to g^a + \tilde{a}$ | $-2t|T^a_{ji}|^2$ |
| F | $\tilde{g}^a + \tilde{g}^b \to \tilde{g}^c + \tilde{a}$ | $-8\frac{(s^2+st+t^2)^2}{st(s+t)}|f^{abc}|^2$ |
| G | $q_i + \tilde{g}^a \to q_j + \tilde{a}$ | $-4(s + \frac{s^2}{t})|T^a_{ji}|^2$ |
| H | $\tilde{q}_i + \tilde{g}^a \to \tilde{q}_j + \tilde{a}$ | $-2(\frac{t}{2} + 2s + 2\frac{s^2}{t})|T^a_{ji}|^2$ |
| I | $q_i + \bar{q}_j \to \tilde{g}^a + \tilde{a}$ | $-4(t + \frac{t^2}{s})|T^a_{ji}|^2$ |
| J | $\tilde{q}_i + \bar{\tilde{q}}_j \to \tilde{g}^a + \tilde{a}$ | $2(\frac{s}{2} + 2t + 2\frac{t^2}{s})|T^a_{ji}|^2$ |

Here $f_{BBF}$, $f_{BFB}$, and $f_{FFF}$ are the products of the corresponding phase space densities,

$$f_{BBF} = f_B(E_1)f_B(E_2)[1 - f_F(E_3)] \,, \tag{3.3}$$
$$f_{BFB} = f_B(E_1)f_F(E_2)[1 + f_B(E_3)] \,, \tag{3.4}$$
$$f_{FFF} = f_F(E_1)f_F(E_2)[1 - f_F(E_3)] \,, \tag{3.5}$$

with
$$f_{B(F)}(E_i) = \frac{1}{\exp(E_i/T) \mp 1} \tag{3.6}$$

since the quarks, gluons, squarks, and gluinos are in thermal equilibrium. The sums of the corresponding squared matrix elements (cf. Table 1) weighted with the appropriate multiplicities and statistical factors read

$$|M_{BBF}|^2 = \frac{g^6(N_c^2-1)}{64\pi^4(f_a/N)^2}\left[\left(s+2t+\frac{2t^2}{s}\right)(N_c+n_f)+\frac{3}{2}s\,n_f\right], \tag{3.7}$$

$$|M_{BFB}|^2 = \frac{g^6(N_c^2-1)}{32\pi^4(f_a/N)^2}\left[\left(-t-2s-\frac{2s^2}{t}\right)(N_c+n_f)-\frac{3}{2}t\,n_f\right], \tag{3.8}$$

$$|M_{FFF}|^2 = \frac{g^6(N_c^2-1)}{32\pi^4(f_a/N)^2}\left[\left(-t-2s-\frac{s^2}{t}+\frac{s^2}{t+s}-\frac{t^2}{s}\right)(N_c+n_f)-\left(\frac{s^2}{t}+\frac{s^2}{t+s}\right)n_f\right], \tag{3.9}$$

where $N_c$ is the number of colors and $n_f$ is the number of color triplet and anti-triplet chiral multiplets.

Since $s = -t - u$, one can write $s + 2t = t - u$ and

$$\pm\frac{s^2}{t}+\frac{s^2}{s+t} = \pm\frac{s^2}{t}-\frac{s^2}{u}\,. \tag{3.10}$$

The difference $t-u$ and $1/t - 1/u$ is odd under exchanging $P_1$ and $P_2$. If the remaining integrand and the measure is even under this transformation, the integral over such terms will be zero. Therefore, the contribution of $s+2t$ in $|M_{BBF}|^2$ will give zero. Moreover, we can substitute $s$ by $-2t$ in $|M_{BBF}|^2$ and use the replacements

$$\frac{s^2}{t}+\frac{s^2}{s+t}\to 0, \qquad -\frac{s^2}{t}+\frac{s^2}{s+t}\to -\frac{2s^2}{t} \tag{3.11}$$

in $|M_{FFF}|^2$. Accordingly,

$$|M_{BBF}|^2 \to \frac{g^6(N_c^2-1)}{32\pi^4(f_a/N)^2}\left[|M_3|^3\,(N_c+n_f)-\frac{3}{2}|M_2|^2 n_f\right], \tag{3.12}$$

$$|M_{BFB}|^2 = \frac{g^6(N_c^2-1)}{32\pi^4(f_a/N)^2}\left[|M_1|^2\,(N_c+n_f)-\frac{3}{2}|M_2|^2 n_f\right], \tag{3.13}$$

$$|M_{FFF}|^2 \to \frac{g^6(N_c^2-1)}{32\pi^4(f_a/N)^2}\left(|M_1|^2-|M_3|^2\right)(N_c+n_f), \tag{3.14}$$

with
$$|M_1|^2 = -t-2s-\frac{2s^2}{t}, \tag{3.15}$$
$$|M_2|^2 = t, \tag{3.16}$$
$$|M_3|^2 = \frac{t^2}{s}\,. \tag{3.17}$$

Only $|M_1|^2$ contributes to the logarithmic cutoff dependence, while the other two squared matrix elements, $|M_2|^2$ and $|M_3|^2$, give finite contributions even for $k_{\rm cut} = 0$. The logarithmic dependence on $k_{\rm cut}$ can be extracted analytically by a partial integration. In the remaining part, $k_{\rm cut}$ can be set to zero. Using the methods described in detail in [26], one obtains the following result for the hard contribution to the axino production rate

$$\left.\frac{d\Gamma_{\tilde{a}}}{d^3 p}\right|_{\rm hard} = \frac{g^6 (N_c^2 - 1)}{32\,\pi^4\,(f_a/N)^2} \bigg[(N_c + n_f)$$
$$\times \bigg\{\frac{T^3 f_F(E)}{128\pi^4}\left[\ln\left(\frac{2^{1/3}T}{k_{\rm cut}}\right) + \frac{11}{4} - \gamma + \frac{\zeta'(2)}{\zeta(2)} + \frac{{\rm Li}_2\left(-e^{-E/T}\right)}{2\pi^2}\right]$$
$$+ I_{BFB}^{|M_1|^2\,{\rm partial}} + I_{FFF}^{|M_1|^2\,{\rm partial}} + I_{BBF}^{|M_3|^2} - I_{FFF}^{|M_3|^2}\bigg\}$$
$$-\frac{3}{2} n_f \left(I_{BBF}^{|M_2|^2} + I_{BFB}^{|M_2|^2}\right)\bigg] \qquad (3.18)$$

where $\gamma = 0.57722$ is Euler's constant, $\zeta(z)$ is Riemann's zeta function with $\zeta'(2)/\zeta(2) = -0.56996$, ${\rm Li}_2(x)$ is the dilogarithm

$$\text{Li}_2(x) = -\int_0^x dt \frac{\ln(1-t)}{t}, \qquad (3.19)$$

and the following integrals appear, which can be evaluated numerically:

$$I_{BFB}^{|M_1|^2\,{\rm partial}} = \frac{1}{256\pi^6 E^2}\int_0^\infty dE_3 \int_0^{E+E_3} dE_1 \ln\left(\frac{|E_1 - E_3|}{E_3}\right) f_{BFB}$$
$$\times \bigg\{\Theta(E - E_1)\left[2E_1 E_2^2 - 2E_1^2 E_2 - (E_1^2 E_2^2 - E_3^2 E^2)\frac{f_B(E_1) + f_F(E_2)}{T}\right]$$
$$-\Theta(E_1 - E_3)\left[2E_1 E_2^2 - 2(E_1^2 + E_3^2)E_2 - E_2^2(E_1^2 + E_3^2)\frac{f_B(E_1) + f_F(E_2)}{T}\right]$$
$$+\Theta(E_3 - E_1)\left[2E_1 E^2 - E^2(E_1^2 + E_3^2)\frac{f_B(E_1) + f_F(E_2)}{T}\right]\bigg\}, \qquad (3.20)$$

$$I_{FFF}^{|M_1|^2\,{\rm partial}} = \frac{1}{256\pi^6 E^2}\int_0^\infty dE_3 \int_0^{E+E_3} dE_1 \ln\left(\frac{|E_1 - E_3|}{E_3}\right) f_{FFF}$$
$$\times \bigg\{\Theta(E - E_1)\left[2E_1 E_2^2 - 2E_1^2 E_2 + \left(E_1^2 E_2^2 - E_3^2 E^2\right)\frac{f_F(E_1) - f_F(E_2)}{T}\right]$$
$$-\Theta(E_1 - E_3)\left[2E_1 E_2^2 - 2(E_1^2 + E_3^2)E_2 + E_2^2(E_1^2 + E_3^2)\frac{f_F(E_1) - f_F(E_2)}{T}\right]$$
$$+\Theta(E_3 - E_1)\left[2E_1 E^2 + E^2(E_1^2 + E_3^2)\frac{f_F(E_1) - f_F(E_2)}{T}\right]\bigg\}, \qquad (3.21)$$

$$I_{BBF(FFF)}^{|M_3|^2} = \frac{1}{(2\pi)^3 2E}\int \left[\prod_{i=1}^3 \frac{d^3 p_i}{(2\pi)^3 2E_i}\right] (2\pi)^4 \delta^4(P_1 + P_2 - P - P_3) f_{BBF(FFF)} |M_3|^2$$

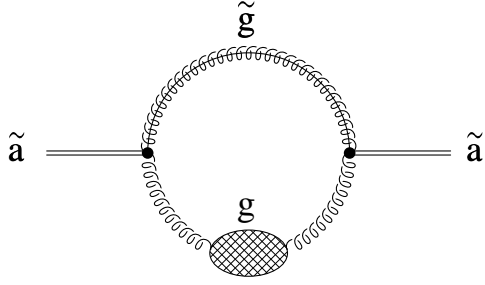

**Figure 2:** The leading contribution to the axino self-energy for soft gluon momentum transfer and hard axino energy. The blob on the gluon line denotes the HTL-resummed gluon propagator.

$$= \frac{1}{256\pi^6 E^2} \int_0^\infty dE_3 \int_0^{E+E_3} dE_2 f_{BBF(FFF)} \left\{ \frac{E_2^2 E^2}{E+E_3} \right.$$
$$\left. +\Theta(E_2 - E_3)(E_3 - E_2)[E_3(E_3 - E_2) + E(E_3 + E_2)] \right\}, \tag{3.22}$$

and

$$I_{BBF(BFB)}^{|M_2|^2} = \frac{1}{(2\pi)^3 2E} \int \left[ \prod_{i=1}^3 \frac{d^3 p_i}{(2\pi)^3 2E_i} \right] (2\pi)^4 \delta^4(P_1 + P_2 - P - P_3) f_{BBF(BFB)} |M_2|^2$$
$$= \frac{1}{768\pi^6 E^2} \int_0^\infty dE_3 \int_0^{E+E_3} dE_1 f_{BBF(BFB)}$$
$$\times \left\{ \Theta(E - E_1)(E - E_1)\left[2E^2 + (3E_3 - E_1)(E + E_1)\right] \right.$$
$$-\Theta(E_1 - E_3) E_2^2 (2E - E_3 + E_1)$$
$$\left. +\Theta(E_3 - E_1)(-3E_3 + 3E_1 - 2E)E^2 \right\}. \tag{3.23}$$

Here it is understood that $E_2 = E + E_3 - E_1$ in (3.20), (3.21), and the second expression in (3.23), while $E_1 = E + E_3 - E_2$ in (3.22).

### 3.2 Soft contribution

We now turn to the contribution from processes involving the exchange of a soft gluon. We compute this soft contribution to the thermal axino production rate from the imaginary part of the thermal axino self-energy [39, 40]

$$\left. \frac{d\Gamma_{\tilde{a}}}{d^3 p} \right|_{\text{soft}} = -\frac{1}{(2\pi)^3 E} f_F(E) \, \text{Im}\Sigma(E + i\epsilon, \mathbf{p})|_{k < k_{\text{cut}}} . \tag{3.24}$$

Concentrating again on the axino-gluino-gluon interaction, the leading order contribution to $\text{Im}\Sigma$ for soft gluon momentum transfer $k < k_{\text{cut}}$ and hard axino energy $E \gtrsim T$ comes from the Feynman diagram shown in Fig. 2. As the gluon momentum transfer is soft, the effective HTL-resummed gluon propagator—indicated by the blob in Fig. 2—has to be used [41, 42]; see also Appendix A. As the hard momentum of the axino flows through the gluino line, there is no need to use a HTL-resummed gluino propagator in the computation of the leading order result.

The effective HTL-resummed gluon propagator takes into account medium effects in the QGSGP such as Debye screening in the static limit [40]. In particular, the Debye screening length $\lambda_D$ defines the thermal gluon mass, $m_g = (\sqrt{3}\lambda_D)^{-1}$. The supersymmetric thermal gluon mass for $N_c$ colors and $n_f$ color triplet and anti-triplet chiral multiplets reads

$$m_g = gT\sqrt{\frac{N_c + n_f}{6}} \ . \tag{3.25}$$

It is derived from the gluon self-energy tensor at finite temperature by taking into account the contributions from gluons, gluinos, quarks, and squarks.

The calculation of the soft contribution to the thermal axino production rate is completely analogous to the axion and gravitino case discussed in detail in [26]: Using the HTL-resummed gluon propagator with the supersymmetric thermal gluon mass (3.25), the axino self-energy can be evaluated in the imaginary-time formalism for discrete imaginary axino energies $p_0 = 2\pi i n T$. Here spectral representations for the propagators can be introduced. Then the sum over the discrete imaginary values of the loop energy $k_0$ is performed. After the analytic continuation of the axino self-energy function $\Sigma(P)$ from the discrete imaginary values $p_0$ to the continuous real axino energy $E$ [39], the imaginary part of $\Sigma(E + i\epsilon, \mathbf{p})$ reduces to an integral over the momentum $\mathbf{k}$ and the spectral parameter $\omega$ of the soft gluon. Accordingly, the soft contribution to the thermal axino production rate (3.24) reduces to the following expression in the high-temperature limit, $T \gg m_{\tilde{q}}, m_{\tilde{g}}$,

$$\left.\frac{d\Gamma^{\tilde{a}}}{d^3 p}\right|_{\text{soft}} = f_F(E) \frac{g^4(N_c^2 - 1)T}{2048\pi^8 (f_a/N)^2}$$
$$\times \int_0^{k_{\text{cut}}} dk\, k^3 \int_{-k}^{k} \frac{d\omega}{\omega} \left[\rho_L(\omega, k)\left(1 - \frac{\omega^2}{k^2}\right) + \rho_T(\omega, k)\left(1 - \frac{\omega^2}{k^2}\right)^2\right], \tag{3.26}$$

where $f_B(\omega) \simeq T/\omega$ has been used as $\omega$ is of order $gT$. For $|\omega| < k$, the spectral densities $\rho_{T/L}$ are given by [43]

$$\rho_T(\omega, k) = \frac{3}{4m_g^2} \frac{x}{(1-x^2)\{[A_T(x)]^2 + [z + B_T(x)]^2\}} \ ,$$
$$\rho_L(\omega, k) = \frac{3}{4m_g^2} \frac{2x}{[A_L(x)]^2 + [z + B_L(x)]^2} \ , \tag{3.27}$$

where $x = \omega/k$, $z = k^2/m_g^2$, and

$$A_T(x) = \frac{3}{4}\pi x, \qquad B_T(x) = \frac{3}{4}\left(2\frac{x^2}{1-x^2} + x\ln\frac{1+x}{1-x}\right) \ ,$$
$$A_L(x) = \frac{3}{2}\pi x, \qquad B_L(x) = \frac{3}{2}\left(2 - x\ln\frac{1+x}{1-x}\right) \ . \tag{3.28}$$

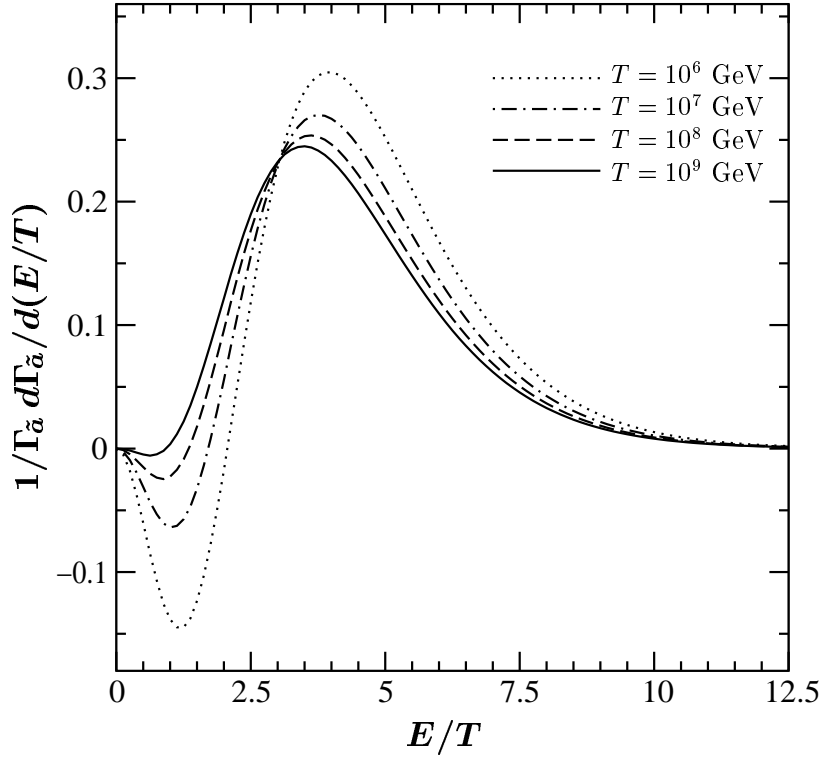

**Figure 3:** The normalized thermal axino production rate $1/\Gamma_{\tilde{a}}\, d\Gamma_{\tilde{a}}/d(E/T)$ as a function of $E/T$ for $N_c = 3$, $n_f = 6$, and temperatures of $T = 10^6\,\text{GeV}$ (dotted line), $T = 10^7\,\text{GeV}$ (dash-dotted line), $T = 10^8\,\text{GeV}$ (dashed line), and $T = 10^9\,\text{GeV}$ (solid line). The results are obtained from (3.30) derived for axino energies $E \gtrsim T$.

For completeness, the relation between the spectral densities and the HTL-resummed gluon propagator is given in Appendix A.

Evaluating the integrals as discussed in [25], one finds

$$\left.\frac{d\Gamma^{\tilde{a}}}{d^3p}\right|_{\text{soft}} = f_F(E)\frac{3g^4(N_c^2-1)m_g^2 T}{4096\pi^8(f_a/N)^2}\left[\ln\left(\frac{k_{\text{cut}}^2}{m_g^2}\right) - 1.379\right]. \tag{3.29}$$

Here the soft singularity is reflected by the logarithm $\ln(k_{\text{cut}}^2/m_g^2)$.

### 3.3 Total thermal production rate

In the sum of the soft and hard contributions the dependence on the unphysical cutoff $k_{\text{cut}}$ cancels and a finite leading order rate is obtained for the production of axinos with $E \gtrsim T$:

$$\frac{d\Gamma_{\tilde{a}}}{d^3p} = \left.\frac{d\Gamma_{\tilde{a}}}{d^3p}\right|_{\text{soft}} + \left.\frac{d\Gamma_{\tilde{a}}}{d^3p}\right|_{\text{hard}}. \tag{3.30}$$

To illustrate this result, we show in Fig. 3 the normalized distribution $1/\Gamma_{\tilde{a}}\, d\Gamma_{\tilde{a}}/d(E/T)$ for $N_c = 3$, $n_f = 6$, and temperatures of $T = 10^6\,\text{GeV}$, $10^7\,\text{GeV}$, $10^8\,\text{GeV}$, and $10^9\,\text{GeV}$. For $E \lesssim T$, we expect modifications as our computation is restricted to the thermal production of hard ($E \gtrsim T$) axinos. The strong coupling is evaluated at the scale $T$ taking into account the one-loop evolution described by the renormalization group equation in the MSSM. Using as input the value of the strong coupling at the Z-boson mass, $g^2(M_Z)/(4\pi) = 0.118$, the running coupling reads

$$g(T) = \left(g^{-2}(M_Z) + \frac{3}{8\pi^2}\ln\left[\frac{T}{M_Z}\right]\right)^{-1/2} , \qquad (3.31)$$

which gives, for example, $g(T = 10^6 \text{ GeV}) = 0.986$ and $g(T = 10^9 \text{ GeV}) = 0.880$. For small values of $E/T \lesssim 2$, the production rate turns negative except for very high temperatures. This unphysical result is due to the extrapolation from the region $g \ll 1$ to $g \approx 1$. Such problems have already been encountered in early applications of the HTL-resummation technique studying, for example, the energy loss of a heavy quark in the quark-gluon plasma [44, 45]. The prescription of Braaten and Yuan discards positive contributions, namely terms in the soft part that do not diverge in the limit $k_{\text{cut}} \to \infty$ and terms in the hard part that do not diverge in the limit $k_{\text{cut}} \to 0$. These terms have to be neglected in the case $g \ll 1$ to obtain a consistent leading order result in $g$. For large values of $E/T$, the Braaten-Yuan method leads to a meaningful result for the energy distribution even though the coupling is close to one. As discussed in detail in the next section, for $T_R \lesssim T_D$, the axino abundance is computed from the integrated production rate

$$\Gamma_{\tilde{a}}(T) = \int_0^\infty d(E/T) \frac{d\Gamma_{\tilde{a}}}{d(E/T)} = \int d^3p \frac{d\Gamma_{\tilde{a}}}{d^3p} . \qquad (3.32)$$

This quantity is positive down to temperatures of about $5 \times 10^3$ GeV. At $T = 10^6$ GeV ($10^7$ GeV), the contribution to $\Gamma_{\tilde{a}}$ from axino energies that give an unphysical rate is about 17% (6%), i.e. hard axinos dominate. Therefore, we will use our result for $\Gamma_{\tilde{a}}$ also for rather small values of $T \sim 10^4$ GeV, keeping in mind that the prediction becomes less reliable for smaller values of $T$. An improved calculation of the thermal axino production rate would require a careful treatment of axinos with $E \lesssim T$ within thermal field theory.

## 4. Relic axino abundance

In this section we compute the relic abundance of axinos from thermal production after completion of the reheating phase of inflation. We work in the scenario in which the axino is the LSP and stable due to $R$-parity conservation. As already discussed in the Introduction, we assume that any initial population of axinos has been diluted away by the exponential expansion during the slow-roll phase of inflation. Only values of the reheating temperature $T_R$ up to the PQ scale $f_a/N$ are considered so that effects from the restoration of the PQ symmetry do not have to be taken into account.

The axino interactions are strongly suppressed by the PQ scale $f_a/N \gtrsim 5 \times 10^9$ GeV; cf. (2.6). Nevertheless, the axinos are in thermal equilibrium for temperatures above a decoupling temperature of $T_D \approx 10^9$ GeV as estimated by Rajagopal, Turner, and Wilczek [16]. Being highly relativistic at such temperatures, the axino number density is given by the equilibrium number density of a relativistic Majorana fermion [19]

$$n_{\tilde{a}}^{\text{eq}} = 3\zeta(3)T^3/(2\pi^2) \,, \tag{4.1}$$

where $\zeta(3) \approx 1.2021$. The axinos decouple from the thermal bath when their interaction rate is approximately equal to the expansion rate of Universe. At high temperatures, this expansion rate is given by the Hubble parameter for the radiation dominated epoch

$$H(T) = \sqrt{\frac{g_*(T)\pi^2}{90}} \frac{T^2}{\text{M}_{\text{Pl}}} \,, \tag{4.2}$$

where $\text{M}_{\text{Pl}} = 2.4 \times 10^{18}$ GeV is the reduced Planck mass. In the MSSM, the number of effectively massless degrees of freedom being in thermal equilibrium with the primordial plasma at the axino decoupling temperature is $g_*(T_D) = 915/4 = 228.75$. This value relies on the assumption of the soft SUSY breaking scale $m_{\text{soft}}$ being significantly below the axino decoupling temperature. Dividing the axino number density $n_{\tilde{a}}$ by the entropy density

$$s(T) = \frac{2\pi^2}{45} g_{*S}(T)T^3 \tag{4.3}$$

with $g_{*S}(T) \approx g_*(T)$ in the radiation dominated epoch, the axino abundance can be expressed conveniently in terms of the yield

$$Y_{\tilde{a}} \equiv \frac{n_{\tilde{a}}}{s} \,, \tag{4.4}$$

which scales out the effect of the expansion of the Universe. For a reheating temperature above the decoupling temperature, $T_R \gtrsim T_D$, the axino yield is governed by the equilibrium number density (4.1) and thus independent of the reheating temperature

$$Y_{\tilde{a}}^{\text{eq}} \equiv \frac{n_{\tilde{a}}^{\text{eq}}}{s} \approx 1.8 \times 10^{-3} \,. \tag{4.5}$$

The contributions from axino production and annihilation processes at smaller temperatures are negligible. Using $h$ $(= 0.71 \pm 0.06$ [3, 9]) to parametrize the Hubble constant $H_0 = 100\,h\,\text{km/sec/Mpc}$, the axino density parameter

$$\Omega_{\tilde{a}}h^2 = \rho_{\tilde{a}}h^2/\rho_c = m_{\tilde{a}}n_{\tilde{a}}h^2/\rho_c = m_{\tilde{a}}Y_{\tilde{a}}s(T_0)h^2/\rho_c \tag{4.6}$$

can be obtained directly from (4.5)

$$\Omega_{\tilde{a}}^{\text{eq}}h^2 \approx \frac{m_{\tilde{a}}}{2\,\text{keV}} \,. \tag{4.7}$$

Here $\rho_c/[s(T_0)h^2] = 3.6 \times 10^{-9}$ GeV has been used based on the following values for the critical density, the present temperature, and the present number of effectively massless degrees of freedom, respectively,

$$\rho_c/h^2 = 8.1 \times 10^{-47} \text{ GeV}^4 \,, \tag{4.8}$$
$$T_0 = 2.73\,K \equiv 2.35 \times 10^{-13} \text{ GeV} \,, \tag{4.9}$$
$$g_{*S}(T_0) = 3.91 \,. \tag{4.10}$$

For a reheating temperature sufficiently below the decoupling temperature, $T_R < T_D$, the axinos are far from being in thermal equilibrium. Thus, the phase space density of axinos $f_{\tilde{a}}$ is negligible in comparison to the phase space densities of the particles in thermal equilibrium. In particular, with the axino number density $n_{\tilde{a}}$ being much smaller than the photon number density $n_\gamma$, the evolution of $n_{\tilde{a}}$ with cosmic time $t$ can be described by the Boltzmann equation

$$\frac{dn_{\tilde{a}}}{dt} + 3H n_{\tilde{a}} = C_{\tilde{a}} \ . \tag{4.11}$$

The second term on the left-hand side of (4.11) accounts for the dilution of the axinos due to the expansion of the Universe described by the Hubble parameter $H$. The collision term $C_{\tilde{a}}$ describes both the production and disappearance of axinos in thermal reactions within the primordial plasma. We can neglect the term for the axino disappearance as it is proportional to $f_{\tilde{a}}$. With the term for the axino production being proportional to $(1 - f_{\tilde{a}}) \approx 1$, the collision term becomes independent of the axino number density. By numerical integration—cf. (3.32)—of the thermal axino production rate (3.30) computed in the previous section, we then obtain the following result for the collision term

$$C_{\tilde{a}}(T)\Big|_{T_R < T_D} \approx \Gamma_{\tilde{a}}(T) = \frac{(N_c^2 - 1)}{(f_a/N)^2} \frac{3\zeta(3) g^6 T^6}{4096 \pi^7} \left[ \ln\left(\frac{1.380\, T^2}{m_g^2}\right)(N_c + n_f) + 0.4336\, n_f \right] . \tag{4.12}$$

This result allows us to calculate the axino abundance from thermal SUSY QCD processes to leading order in the gauge coupling $g$, contrary to previous estimates, which depend on an unknown scale of the logarithmic term [14].

The Boltzmann equation can now be solved analytically using standard arguments [19]. Assuming the conservation of entropy per comoving volume, $sR^3 = \text{const.}$, where $R$ is the scale factor of the Universe, the Boltzmann equation (4.11) can be rewritten in terms of the yield

$$\frac{d}{dt} Y_{\tilde{a}} = \frac{C_{\tilde{a}}}{s} \ . \tag{4.13}$$

As the thermal axino production proceeds basically during the hot radiation dominated epoch, i.e. at temperatures above the one at matter-radiation equality ($T_{\text{mat=rad}}$), we can change variables from cosmic time to temperature, $dt = -dT/[H(T)T]$, with $H(T)$ given in (4.2). After separating variables, the Boltzmann equation (4.13) can be integrated. The resulting axino yield at the present temperature of the Universe $T_0$ is given by

$$Y_{\tilde{a}}(T_0) \approx Y_{\tilde{a}}(T_{\text{mat=rad}}) - Y_{\tilde{a}}(T_R) = \int_{T_{\text{mat=rad}}}^{T_R} dT \frac{C_{\tilde{a}}(T)}{T s(T) H(T)} \ , \tag{4.14}$$

where $Y_{\tilde{a}}(T_R) = 0$ as we consider only axino production after the completion of the reheating phase. Since the factor $T^6$ in the collision term (4.12) cancels against a corresponding factor in the denominator of the integrand in (4.14), we can neglect the temperature dependence for the integration and find

$$Y_{\tilde{a}}(T_0) \approx \frac{C_{\tilde{a}}(T_R)}{s(T_R) H(T_R)} = 2.0 \times 10^{-7} g^6 \ln\left(\frac{1.108}{g}\right) \left(\frac{10^{11}\,\text{GeV}}{f_a/N}\right)^2 \left(\frac{T_R}{10^4\,\text{GeV}}\right) , \tag{4.15}$$

where we have used $N_c = 3$, $n_f = 6$, and $g_{*S}(T_R) = 228.75$. Inserting (4.15) into (4.6), the corresponding axino density parameter is obtained

$$\Omega_{\tilde{a}} h^2 = 5.5 \, g^6 \ln\left(\frac{1.108}{g}\right) \left(\frac{m_{\tilde{a}}}{0.1\,\mathrm{GeV}}\right) \left(\frac{10^{11}\,\mathrm{GeV}}{f_a/N}\right)^2 \left(\frac{T_R}{10^4\,\mathrm{GeV}}\right) . \qquad (4.16)$$

The yield (4.15) and the density parameter (4.16) become negative for values of the strong coupling $g \gtrsim 1.11$. This is not surprising since, as already mentioned in Sec. 3.3, the HTL resummation requires the separation of scales $gT \ll T$ which implies $g \ll 1$. By using (4.15) and (4.16) with the running strong coupling $g(T_R)$ given in (3.31), we find that our result for the relic axino abundance can only be meaningful for $T_R \gtrsim 10^4$ GeV. Towards lower values of the reheating temperature, the effect of higher orders will become more important. Moreover, also the finite masses of the particles and axino production by gluino decay cannot be neglected for reheating temperatures towards $10^4$ GeV. In particular, the number of effective relativistic degrees of freedom $g_{*S}(T_R)$ becomes smaller for a reheating temperature close to the soft SUSY breaking scale $m_{\mathrm{soft}}$. Since $g_{*S}(T_R) = 106.75$ if only the full set of particles in the minimal standard model contributes to the relativistic degrees of freedom, the reduction of the effectively massless degrees of freedom could enhance the yield at most by about a factor of two.

In Fig. 4 our result (4.15) for the axino yield $Y_{\tilde{a}}$ is illustrated as a function of the reheating temperature $T_R$ for a PQ scale of $f_a/N = 10^{10}$ GeV (dash-dotted line), $10^{11}$ GeV (solid line), and $10^{12}$ GeV (dashed line). Here the 1-loop running of the strong coupling in the MSSM (3.31) is taken into account by replacing $g$ with $g(T_R)$ in (4.15). The yield obtained with the HTL-resummation technique is about one order of magnitude below the one reported in Ref. [14], which was obtained by regularizing the IR singularities by putting in by hand an effective gluon mass $m_{\mathrm{eff}} = gT$. In contrast, the simple estimates of Asaka and Yanagida [12] agree amazingly well with our result. As the yield (4.15) approaches the equilibrium value $Y_{\tilde{a}}^{\mathrm{eq}}$ given in (4.5), the axino disappearance processes should be taken into account. This would lead to a smooth approach of the non-equilibrium yield to the equilibrium abundance. Without the backreactions taken into account, the kink position indicates a lower bound for the corresponding axino decoupling temperature $T_D$, which agrees with the estimate of $T_D$ in [16]. For example, for $f_a/N = 10^{11}$ GeV, we find that the axino decoupling temperature is at least $10^9$ GeV. Towards reheating temperatures of $10^4$ GeV, we have to stress again that the uncertainty in the yield increases due to the theoretical uncertainties in the thermal production rate discussed in Sec. 3.3.

In Fig. 5 our result (4.16) for the relic axino density $\Omega_{\tilde{a}} h^2$ is illustrated as a function of the reheating temperature $T_R$ for a PQ scale of $f_a/N = 10^{11}$ GeV and an axino mass of $m_{\tilde{a}} = 1\,\mathrm{keV}$ (dotted line), $100\,\mathrm{keV}$ (dashed line), and $10\,\mathrm{MeV}$ (solid line). The 1-loop running of the strong coupling in the MSSM (3.31) is again taken into account by replacing $g$ with $g(T_R)$ in (4.16). For a reheating temperature $T_R$ above the axino decoupling temperature $T_D$ of about $10^9$ GeV, the relic axino density (4.7) is independent of $T_R$ due to the thermal equilibrium as shown for $m_{\tilde{a}} = 1\,\mathrm{keV}$ by the horizontal line. Again there will be a smooth transition instead of a kink once the axino disappearance processes are taken into account. The grey band in Fig. 5 indicates the WMAP result on the cold dark matter density ($2\sigma$ error) $\Omega_{\text{CDM}}^{\text{WMAP}} h^2 = 0.113^{+0.016}_{-0.018}$ [9].

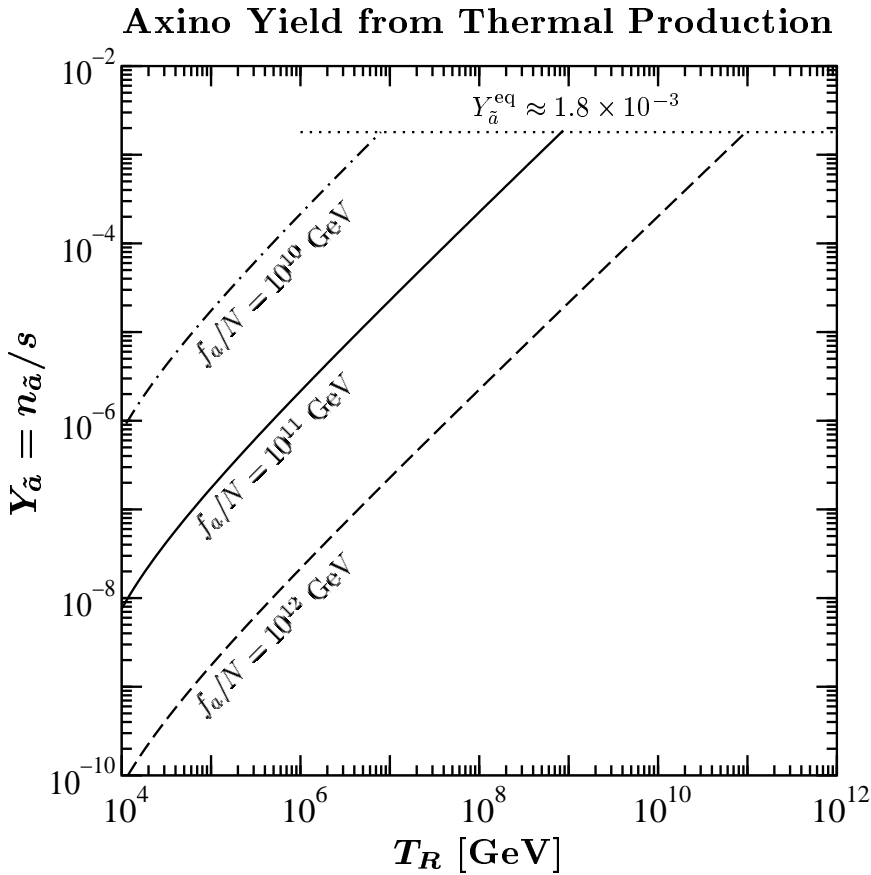

**Figure 4:** The axino yield, $Y_{\tilde{a}} = n_{\tilde{a}}/s$, from thermal production in the QGSGP as a function of the reheating temperature $T_R$ for a PQ scale of $f_a/N = 10^{10}\,\text{GeV}$ (dash-dotted line), $10^{11}\,\text{GeV}$ (solid line), and $10^{12}\,\text{GeV}$ (dashed line). For $T_R$ above the axino decoupling temperature $T_D$, the axinos have been in thermal equilibrium with the primordial plasma. The resulting yield $Y_{\tilde{a}}^{\text{eq}}$ is independent of $T_R$ as indicated by the dotted line.

## 5. Axino dark matter

In this section we study the cosmological implications of the relic axino abundance computed in the previous section. The possibility of axinos forming a dominant part of cold dark matter is discussed. We also address briefly the nucleosynthesis constraints taking into account results from recent studies of potential NLSP decays [18, 14, 11].

If axinos are stable, they contribute to the present mass density of dark matter in the Universe. For the corresponding relic axino abundance from thermal production after completion of the reheating phase of inflation, we have obtained the results (4.7) and (4.16) illustrated in Fig. 5. Accordingly, the amount of axino dark matter depends on the axino mass $m_{\tilde{a}}$ and the reheating temperature $T_R$ once the PQ scale $f_a/N$ is fixed. As one can see in Fig. 5, there are certain combinations of $m_{\tilde{a}}$ and $T_R$ for which the relic axino abundance agrees with the WMAP result on the cold dark matter density. Thus, as far as the abundance is concerned, axinos could provide the dominant part of dark matter.

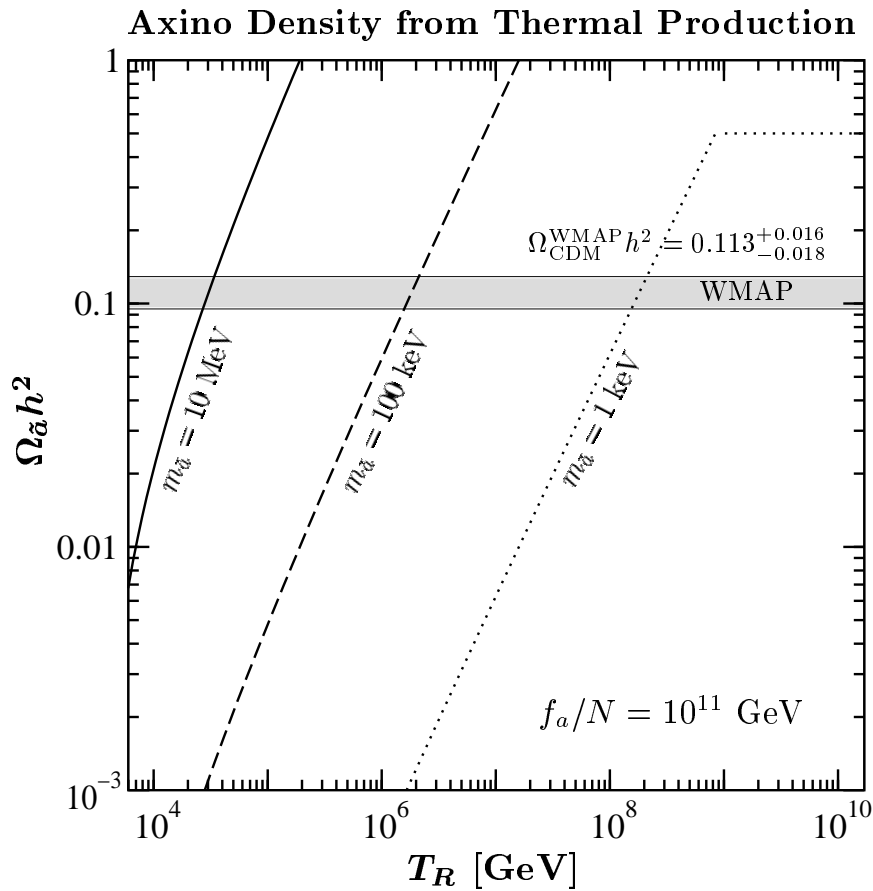

**Figure 5:** The axino density parameter $\Omega_{\tilde{a}}h^2$ as a function of the reheating temperature $T_R$ for a PQ scale of $f_a/N = 10^{11}\,\text{GeV}$ and an axino mass of $m_{\tilde{a}} = 1\,\text{keV}$ (dotted line), $100\,\text{keV}$ (dashed line), and $10\,\text{MeV}$ (solid line). For $T_R$ above the axino decoupling temperature $T_D$ of about $10^9\,\text{GeV}$, the axinos have been in thermal equilibrium with the primordial plasma. The resulting density $\Omega_{\tilde{a}}^{\text{eq}}h^2$ is independent of $T_R$ as shown for $m_{\tilde{a}} = 1\,\text{keV}$. The grey band indicates the WMAP result on the cold dark matter density ($2\sigma$ error) $\Omega_{\text{CDM}}^{\text{WMAP}}h^2 = 0.113^{+0.016}_{-0.018}$ [9].

The next question is whether axino dark matter from thermal production is sufficiently cold (i.e. non-relativistic sufficiently long before matter-radiation equality) in order to explain the formation and power spectrum of large-scale structures and the early reionization observed by the WMAP satellite. For reheating temperatures $T_R$ below the axino decoupling temperature $T_D$, axinos are not in thermal equilibrium. However, as the thermal production proceeds (by definition) through reactions of particles in thermal equilibrium, the axinos are produced in kinetic equilibrium with the thermal plasma, i.e. the axino production rate shows basically a thermal spectrum (cf. Sec. 3.3). In particular, the axinos become non-relativistic when the temperature of the QGSGP is of order of the axino mass, $T \sim m_{\tilde{a}}$. We thus can adopt also for $T_R \lesssim T_D$ the dark matter classification used for particles with thermal spectrum, where only the mass of the dark matter particle is relevant to decide whether the dark matter is cold, warm, or hot. For an axino mass in the range $m_{\tilde{a}} \lesssim 1\,\text{keV}$, $1\,\text{keV} \lesssim m_{\tilde{a}} \lesssim 100\,\text{keV}$, and $m_{\tilde{a}} \gtrsim 100\,\text{keV}$, we refer to hot, warm, and cold axino dark matter, respectively. This classification can be regarded only as an approximate guideline. In the literature, for example, also particles with masses of about $0.2\,\text{keV}$ have been considered as warm dark matter [46]. For our investigation, an important bound results from simulations of early structure formation, which show that dark matter of particles with mass $m_x \lesssim 10\,\text{keV}$ cannot explain the early reionization observed by the WMAP satellite; cf. [47] and references therein.

Assuming a PQ scale of $f_a/N = 10^{11}\,\text{GeV}$, Fig. 5 shows that thermally produced axinos could provide the dominant part of *cold* dark matter for a reheating temperature of $T_R \lesssim 10^6\,\text{GeV}$. For a higher reheating temperature, the mass of the LSP axino has to be smaller. Otherwise, the relic axino abundance would exceed the WMAP result for the relic dark matter abundance. According to our classification, axinos would be *warm* dark matter up to a reheating temperature of about $10^8\,\text{GeV}$ and *hot* dark matter beyond this value. For an axino mass of $m_{\tilde{a}} \lesssim 10\,\text{keV}$ or a reheating temperature of $T_R \gtrsim 10^7\,\text{GeV}$, the simulations mentioned above [47] tell us that an additional amount of *colder* dark matter is needed to explain the structures required for the early reionization of the Universe.

Let us now discuss the constraint concerning a potential increase of the energy density at the time of primordial nucleosynthesis. The production of energetic axinos shortly before nucleosynthesis could lead to such an increase [14], which would speed up the expansion of the Universe. As already mentioned in the Introduction, the resulting earlier freeze out of species would then change the ratio of protons to neutrons and thus the abundance of the light elements (D, $^3$He, $^4$He, Li). Concentrating on reheating temperatures $T_R \gtrsim 10^4\,\text{GeV}$, the significant amount of axinos is produced thermally at $T_R$ and thus long before nucleosynthesis, which takes place around $T = 1\,\text{MeV}$. For $m_{\tilde{a}} \gtrsim 1\,\text{MeV}$, the axinos are non-relativistic at the time of primordial nucleosynthesis. Accordingly, their contribution to the energy density becomes important only at matter-radiation equality. For $m_{\tilde{a}} \lesssim 1\,\text{MeV}$, the axinos are still relativistic at $T \approx 1\,\text{MeV}$. As their spectrum follows basically a thermal distribution also for $T_R < T_D$, their energy density can be approximated by $\rho_{\tilde{a}} \approx 7\pi^2 T_{\tilde{a}}^4/120$, where the effective axino temperature $T_{\tilde{a}}$ at the time of primordial nucleosynthesis (BBN) is much lower than the corresponding neutrino temperature,

$$\left(\frac{T_{\tilde{a}}}{T_\nu}\right)_{\text{BBN}} = \left(\frac{g^*(1\,\text{MeV})}{g^*(\min\{T_R, T_D\})}\right)^{1/3} = \left(\frac{10.75}{228.75}\right)^{1/3}. \qquad (5.1)$$

Therefore, the axinos contribute much less to the energy density at $T \approx 1\,\text{MeV}$ than one light neutrino species,

$$\left(\frac{\rho_{\tilde{a}}}{\rho_\nu}\right)_{\text{BBN}} = \left(\frac{10.75}{228.75}\right)^{4/3} \approx 0.017. \qquad (5.2)$$

Such a small increase of the energy density at the time of primordial nucleosynthesis is consistent with the observed abundance of the light elements.

More model dependent are the nucleosynthesis constraints concerning the potential destruction of the light primordial elements by photons or hadronic showers from late decays of the NLSP into axinos. Here the constraints depend very much on the properties of the NLSP, i.e. its composition, its mass relative to the axino mass $m_{\tilde{a}}$, and its coupling to the axino. The situations in which the NLSP mass is close to the axino mass $m_{\tilde{a}}$ and/or in which the coupling to the axino is particularly small are most dangerous since this decreases the width of the corresponding decay. If the decays occur during or after primordial nucleosynthesis, the photons or hadronic showers accompanying the axino can destroy efficiently the produced light elements. With the axino being the LSP, two attractive candidates for the NLSP are a bino-dominated lightest neutralino $\chi$ and the lighter stau $\tilde{\tau}$. In the case of a bino-dominated lightest neutralino $\chi$ NLSP, the decays $\chi \to \tilde{a}\gamma$, $\chi \to \tilde{a}\gamma^* \to \tilde{a}q\bar{q}$, and $\chi \to \tilde{a}Z/Z^* \to \tilde{a}q\bar{q}$ provide photons and hadronic showers, which could destroy the light elements. In the case of a lighter stau $\tilde{\tau}$ NLSP, the decays $\tilde{\tau} \to \tilde{a}\tau \to \tilde{a}q\bar{q}'\nu_\tau$ and $\tilde{\tau} \to \tilde{a}\tau\gamma/Z \to \tilde{a}q\bar{q}'\nu_\tau\gamma/Z$ are potentially dangerous. As the real photons will thermalize on the background electrons, positrons, and photons (including those in the high-energy tail of the cosmic background radiation), the decay lifetime into photons can be as large as $10^4$ seconds without being problematic. Therefore, the constraints from decays into hadronic showers are more severe. Both NLSP scenarios have been studied carefully in [18, 14, 11], where explicit expressions for the corresponding decay widths can be found. Because of the dependence on several unknown quantities, we will not go into more details but emphasize that additional bounds on the axino mass might arise from these nucleosynthesis constraints. As the axino interactions are not as strongly suppressed as the gravitino interactions ($f_a/N \ll \mathrm{M}_{\mathrm{Pl}}$), the axino LSP anyhow is far less problematic than the gravitino LSP with respect to these nucleosynthesis constraints. Indeed, a low-energy spectrum, in which the axino is the LSP and the gravitino the NLSP, has been suggested as a solution [12] to the gravitino problem [13], i.e. to avoid cosmologically dangerous decays.

We assume from now on that the nucleosynthesis constraints are satisfied for any value of the axino mass up to 100 MeV. Thermally produced axinos then provide the dominant part of dark matter for the combinations of the axino mass $m_{\tilde{a}}$ and the reheating temperature $T_R$ indicated by the grey band in Fig. 6. For the $m_{\tilde{a}}$–$T_R$ combinations within this grey band, the relic axino density (obtained for $f_a/N = 10^{11}$ GeV) agrees with the WMAP result ($2\sigma$ errors) on the cold dark matter density. In the scenario in which axinos are the dominant part of dark matter, the WMAP result allows us to extract upper limits on the reheating temperature as a function of the axino mass. Note that these limits are relaxed by one order of magnitude with respect to the limits found in the investigation of Covi et al. [14], where a similar curve has been obtained with the constraint $\Omega_{\tilde{a}}h^2 < 1$. This difference results from our computation of the thermal production rate, in which we have used the Braaten-Yuan prescription with the HTL-resummed gluon propagator as opposed to the pragmatic cutoff procedure used in [14]. For $T_R \gtrsim 10^9$ GeV, the axinos have been in thermal equilibrium with the primordial plasma. As already discussed in the previous section, the resulting relic axino density (4.7) is completely determined by $m_{\tilde{a}}$ and independent of $T_R$. The agreement with the WMAP result is achieved for any $T_R \gtrsim 10^9$ GeV as long as $m_{\tilde{a}} \approx 0.2$ keV, which is the updated version of the Rajagopal-

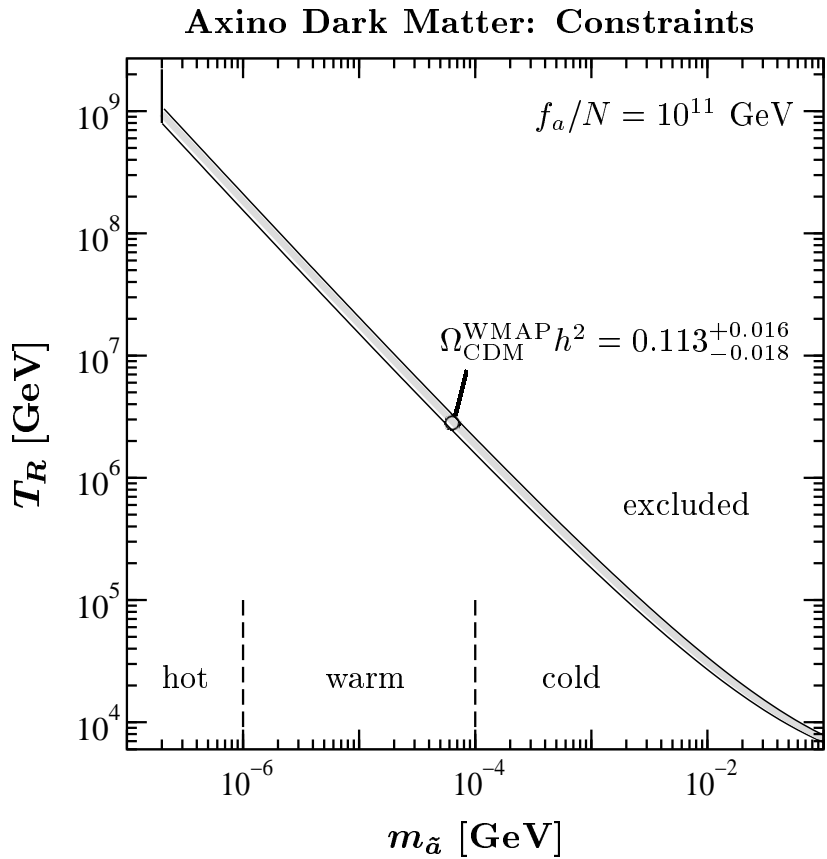

**Figure 6:** Constraints for axino dark matter from thermal production in the early Universe. For the combinations of the axino mass $m_{\tilde{a}}$ and the reheating temperature $T_R$ within grey band, the density parameter of thermally produced axinos $\Omega_{\tilde{a}} h^2$ (obtained for $f_a/N = 10^{11}\,\text{GeV}$) agrees with the WMAP result on the cold dark matter density ($2\sigma$ errors) [9] $\Omega_{\text{CDM}}^{\text{WMAP}} h^2 = 0.113^{+0.016}_{-0.018}$. We refer to hot, warm, and cold axino dark matter for $m_{\tilde{a}} \lesssim 1\,\text{keV}$, $1\,\text{keV} \lesssim m_{\tilde{a}} \lesssim 100\,\text{keV}$, and $m_{\tilde{a}} \gtrsim 100\,\text{keV}$, respectively. For $T_R \gtrsim 10^9\,\text{GeV}$, the axinos have been in thermal equilibrium with the primordial plasma so that the relic axino density is independent of $T_R$. For axinos with $m_{\tilde{a}} \lesssim 10\,\text{keV}$, which cannot explain the structure formation needed for the observed early reionization [47], the WMAP result provides a conservative upper bound on the reheating temperature $T_R$.

Turner-Wilczek bound [16, 14]. Since the more severe upper bounds on the amount of hot and warm dark matter in the Universe are still controversial, we have extended the bound from WMAP as a conservative upper limit also to the regions of warm and hot axino dark matter.

Let us suggest two scenarios to illustrate the possible significance of the upper limits on the reheating temperature $T_R$. (i) If axinos provide the density of cold dark matter observed by WMAP, the reheating temperature after inflation must have been relatively small, $T_R \lesssim 10^6\,\text{GeV}$. Such a low reheating temperature excludes some models for inflation. Moreover, the baryon asymmetry in the Universe has to be explained by a mechanism working efficiently already at relatively small temperatures. (ii) If the baryon asymmetry in the Universe is generated by thermal leptogenesis [27], which requires reheating temperatures of $T_R \gtrsim 10^9$ GeV, axinos can only be hot dark matter. The cold dark matter and large-scale-structure formation could then be explained by the axion as suggested in [34]. The gravitino problem, which one faces for such high reheating temperatures, can be solved as decays of the gravitino into an axion and an axino are possible [12].

## 6. Conclusion and Outlook

If supersymmetry is realized in nature and the strong CP problem is solved by the Peccei-Quinn mechanism, the axino should exist as the fermionic superpartner of the axion. As the interactions of the electrically and color neutral axion supermultiplet are suppressed by the PQ scale $f_a/N \gtrsim 5 \times 10^9$ GeV, the axino couples only very weakly to the MSSM particles. If the axino is the lightest supersymmetric particle and if $R$-parity is conserved, axinos will be stable and exist as dark matter in our Universe. Then, depending on the axino mass $m_{\tilde{a}}$ and the temperature of the primordial plasma, axinos could play an important role in the cosmos. In particular, they could provide the dominant part of cold dark matter.

We have studied the thermal production of axinos in the early Universe. Assuming that inflation has governed the earliest moments of the Universe, any initial population of axinos has been diluted away by the exponential expansion during the slow-roll phase. The thermal production of axinos starts then after completion of the reheating phase at the reheating temperature $T_R$. We have restricted our investigation to $T_R \gtrsim 10^4$ GeV, where axino production from decays of particles out of equilibrium is negligible.

We have computed the thermal production rate of axinos with $E \gtrsim T$ in SUSY QCD to leading order in the gauge coupling. Using the Braaten-Yuan prescription and the HTL-resummed gluon propagator, we have obtained a finite result in a gauge-invariant way, which takes into account Debye screening in the hot QGSGP. In particular, the derived axino production rate depends logarithmically on the supersymmetric gluon plasma mass, which regularizes the infrared divergence occurring in leading order. An important question concerns the size of higher-order corrections. Our computation relies on the separation of scales $gT \ll T$ valid only in the weak coupling limit $g \ll 1$. Since $g(T=10^6 \text{ GeV}) = 0.986$ and $g(T=10^9 \text{ GeV}) = 0.880$ according to the 1-loop running of the strong coupling in the MSSM, the obtained rate is most reliable for temperatures of $T \gtrsim 10^9$, while higher-order corrections might become sizable for temperatures of $T \lesssim 10^6$ GeV. The development of a framework, which allows us to compute reliably thermal production rates also for $g \gtrsim 1$, is a hard conceptual problem of utmost importance, in particular, for quark-gluon plasma physics. The computation of the thermal production rate for axinos with $E \lesssim T$ is another non-trivial future task. At present, the result on the thermal axino production rate given in this work supersedes previous estimates since it does not depend on *ad hoc* cutoffs used to regularize the infrared divergence in earlier studies [14].

Based on our calculation of the thermal axino production rate, we have discussed quantitatively the role of the axino in cosmology. For a reheating temperature $T_R$ below the axino decoupling temperature $T_D$ of about $10^9$ GeV, the axinos have never been in thermal equilibrium. Then, with our restriction to $T_R \gtrsim 10^4$ GeV, the relic axino abundance is governed by the thermal axino production in reactions of colored (SUSY QCD) particles in the primordial MSSM plasma. The evolution of the axino number density is described by the Boltzmann equation with the corresponding collision term. With axino disappearance processes being negligible, we have computed this collision term by integrating our result for the thermal axino production rate. By solving the resulting Boltzmann equation, we have obtained the relic axino density $\Omega_{\tilde{a}}h^2$, which depends on the axino mass, the PQ scale, and the reheating temperature. For certain combinations of these quantities, the derived relic axino density indeed agrees with the WMAP result on the abundance of cold dark matter, $\Omega_{\tilde{a}}h^2 \approx \Omega_{\text{CDM}}^{\text{WMAP}}h^2$. For example, for $f_a/N = 10^{11}$ GeV, this agreement is achieved with $m_{\tilde{a}} = 100$ keV and $T_R \approx 10^6$ GeV. Although relatively light for being cold dark matter, axinos with a mass of $m_{\tilde{a}} = 100$ keV could still explain large-scale-structure formation, the corresponding power spectrum, and the early reionization observed by WMAP. For larger axino masses, the matching with the WMAP result does restrict the reheating temperature to lower values. Already a reheating temperature of $T_R \approx 10^6$ GeV is relatively small and excludes some models for inflation and baryogenesis such as thermal leptogenesis. Nevertheless, there are several baryogenesis scenarios—such as EW baryogenesis within SUSY (cf. [48] and references therein) and leptogenesis via inflaton decay or right-handed sneutrino condensation (cf. [49, 50] and references therein)—which can explain the baryon asymmetry in the Universe for a reheating temperature as low as $10^6$ GeV. The gravitino problem can be avoided for $T_R \approx 10^6$ GeV. Thus, the above combination illustrates that axinos might very well provide the dominant part of cold dark matter.

We should emphasize that our results for the relic axino abundance are an order of magnitude below the ones reported in Ref. [14]. Therefore, the reheating temperature, at which the axino density agrees with the WMAP result on the cold dark matter density, has increased by one order of magnitude for fixed values of $f_a/N$ and $m_{\tilde{a}}$. This effect is due to our computation of the thermal axino production rate, which is based on the HTL-resummation technique as opposed to the more pragmatic cutoff procedure used in [14].

For a reheating temperature above the temperature $T_D \approx 10^9$ GeV at which the axinos decouple from the thermal bath, axinos have been in thermal equilibrium with the primordial plasma. The resulting relic axino abundance, $\Omega_{\tilde{a}}^{\text{eq}}h^2 \approx 0.5\,m_{\tilde{a}}/\text{keV}$, is independent of the reheating temperature and the PQ scale. With the upper limit on the axino density parameter given by the WMAP result on the relic abundance of cold dark matter, one finds $m_{\tilde{a}} \lesssim 0.2$ keV. Accordingly, axinos are too light for being cold dark matter. As warm or hot dark matter, they alone cannot explain the formation and power-spectrum of cosmic large-scale structures and, in particular, the early reionization observed by WMAP. Nevertheless, even such a light axino can have important cosmological implications. By destabilizing the other $R$-odd particles including the gravitino, the light axino provides a solution to the gravitino problem so that a high reheating temperature of $T_R \gtrsim 10^9$ GeV is no longer problematic [12]. Thus, with the light axino being the LSP, thermal leptogenesis [27] can coexist with SUSY and explain the baryon asymmetry in the Universe. In addition, the axion could provide the dominant part of cold dark matter [34].

In the scenarios discussed above, we have assumed that the properties of the NLSP are such that its decays do not affect the abundance of primordial light elements. However, depending on the composition, the mass, and the couplings of the NLSP, this nucleosynthesis constraint could induce additional limits possibly excluding the discussed scenarios. To clarify this issue, experimental insights into the spectrum of superparticles are crucial.

Another intriguing question is whether the axino itself could ever be detected and identified despite of its extremely weak coupling to the MSSM particles. With a stable axino LSP, the NLSP will appear as a stable superparticle in most collider experiments. The decay of the NLSP into axinos will usually take place far outside of the detector and thus will be hard to observe. Nevertheless, for the stable gravitino LSP scenario, it has been demonstrated recently that the gravitino could be identified unambiguously provided the NLSP is a charged slepton such as the lighter stau [51]. This proposal can be carried over to the axino LSP scenario. One would ideally produce the charged NLSP sleptons in an $e^+e^-$ collider such that they stop within the detector. These stopped sleptons will decay into axinos and leptons within the detector at much later times. Although the axino couplings are much more model dependent than the ones of the gravitino, the analysis of such decays could verify the existence of the axino as the stable LSP. Independently of this very specific scenario, the determination of the axino mass will be crucial to decide if axinos are present as cold dark matter in our Universe.

## Acknowledgments

We would like to thank Wilfried Buchmüller, Laura Covi, Koichi Hamaguchi, Arthur Hebecker, Michael Ratz, Andreas Ringwald, Markus Thoma, Uwe-Jens Wiese, and Yvonne Wong for helpful discussions and valuable comments on the manuscript. F.D.S. is grateful to the Institute for Theoretical Physics of the University of Bern for generous hospitality during the completion of parts of this work.

## A. Effective HTL-resummed gluon propagator

In covariant gauge, the HTL-resummed gluon propagator has the form [41, 42, 43]

$$i\Delta_{\mu\nu}(K) = i\left(A_{\mu\nu}\Delta_T + B_{\mu\nu}\Delta_L + C_{\mu\nu}\xi\right) , \tag{A.1}$$

with the tensors

$$A_{\mu\nu} = -g_{\mu\nu} - \frac{1}{k^2}\left[K^2 v_\mu v_\nu - K\cdot v(K_\mu v_\nu + K_\nu v_\mu) + K_\mu K_\nu\right] ,$$

$$B_{\mu\nu} = v_\mu v_\nu - \frac{K\cdot v}{K^2}(K_\mu v_\nu + K_\nu v_\mu) + \left(\frac{K\cdot v}{K^2}\right)^2 K_\mu K_\nu ,$$

$$C_{\mu\nu} = \frac{K_\mu K_\nu}{(K^2)^2} , \tag{A.2}$$

and the transverse $(T)$ and longitudinal $(L)$ propagators

$$\Delta_T(k_0, k) = \frac{1}{k_0^2 - k^2 - \Pi_T(k_0, k)} ,$$

$$\Delta_L(k_0, k) = \frac{1}{k^2 - \Pi_L(k_0, k)} . \tag{A.3}$$

Here $K = (k_0, \mathbf{k})$ with $k_0 = i2\pi nT$ and $k = |\mathbf{k}|$, $\xi$ is a gauge-fixing parameter, $v$ is the velocity of the thermal bath and $\Pi_{T/L}$ are the transverse ($T$) and longitudinal ($L$) self-energies of the gluon. The spectral representation of the propagators $\Delta_{T/L}$ reads

$$\Delta_{T/L}(k_0, k) = \int_{-\infty}^{\infty} d\omega \frac{1}{k_0 - \omega} \rho_{L/T}(\omega, k) \tag{A.4}$$

with spectral densities given in (3.27) and (3.28) in the main text.

# Erratum to: "Axino dark matter from thermal production"
# [J. Cosmol. Astropart. Phys. JCAP08(2004)008]

Arnd Brandenburg[1,a] and Frank Daniel Steffen[2,b]

[1] *Genedata AG, Maulbeerstr. 46, CH-4016 Basel, Switzerland*

[2] *Max-Planck-Institut für Physik, Föhringer Ring 6, D–80805 Munich, Germany*

Equation (3.18) has to be replaced by

$$\left.\frac{d\Gamma_{\tilde{a}}}{d^3 p}\right|_{\text{hard}} = \frac{g^6 (N_c^2 - 1)}{32\pi^4 (f_a/N)^2} \left[ (N_c + n_f) \right. \quad \text{(E.1)}$$

$$\times \left\{ \frac{T^3 f_F(E)}{128\pi^4} \left[ \ln\left(\frac{2^{1/3} T}{k_{\text{cut}}}\right) + \frac{17}{6} - \gamma + \frac{\zeta'(2)}{\zeta(2)} \right] \right.$$

$$\left. + I_{BFB}^{|M_1|^2 \text{ partial}} + I_{FFF}^{|M_1|^2 \text{ partial}} + I_{BBF}^{|M_3|^2} - I_{FFF}^{|M_3|^2} \right\} - \frac{3}{2} n_f \left( I_{BBF}^{|M_2|^2} + I_{BFB}^{|M_2|^2} \right) \right]$$

The difference between Eqs. (3.18) and (E.1) is a non-vanishing surface term in the BFB case which was overlooked in [1]. Accordingly, the numbers in the logarithm of the collision term (4.12) and in the logarithm of the yield (4.15) and the relic density (4.16) change from 1.380 to 1.647 and from 1.108 to 1.211, respectively. The corrected expressions are

$$\left. C_{\tilde{a}}(T) \right|_{T_R < T_D} \approx \Gamma_{\tilde{a}}(T) = \frac{(N_c^2 - 1)}{(f_a/N)^2} \frac{3\zeta(3)g^6 T^6}{4096\pi^7} \left[ \ln\left(\frac{1.647\, T^2}{m_g^2}\right) (N_c + n_f) + 0.4336\, n_f \right], \quad \text{(E.2)}$$

$$Y_{\tilde{a}}(T_0) \approx \frac{C_{\tilde{a}}(T_R)}{s(T_R) H(T_R)} = 2.0 \times 10^{-7} g^6 \ln\left(\frac{1.211}{g}\right) \left(\frac{10^{11}\,\text{GeV}}{f_a/N}\right)^2 \left(\frac{T_R}{10^4\,\text{GeV}}\right), \quad \text{(E.3)}$$

$$\Omega_{\tilde{a}} h^2 = 5.5\, g^6 \ln\left(\frac{1.211}{g}\right) \left(\frac{m_{\tilde{a}}}{0.1\,\text{GeV}}\right) \left(\frac{10^{11}\,\text{GeV}}{f_a/N}\right)^2 \left(\frac{T_R}{10^4\,\text{GeV}}\right). \quad \text{(E.4)}$$

Using the running coupling (3.31) to evaluate $g = g(T_R)$, the axino abundance described by (E.3) and (E.4) exceeds the one described by (4.15) and (4.16) by factors of 1.4, 1.8, and 4.8 for $T_R = 10^9$, $10^6$, and $10^4$ GeV, respectively. The qualitative findings of [1] are not affected and all figures except for Fig. 3 change only mildly in the region in which the applied methods are most reliable, i.e., for $T \gg 10^4$ GeV. The corrected Fig. 3 is shown below with numberings and captions adopted from [1]. The two sentences after (3.32) have to be modified accordingly: "This quantity is positive down to temperatures of about 100 GeV. At $T = 10^6$ GeV ($10^4$ GeV), the contribution to $\Gamma_{\tilde{a}}$ from axino energies that give an unphysical rate is about 0.6% (17%), i.e. hard axinos dominate." Note also that there is a typographical error in (3.12): $|M_3|^3$ should read $|M_3|^2$.

---

[a] Arnd.Brandenburg@genedata.com
[b] Frank.Steffen@mppmu.mpg.de

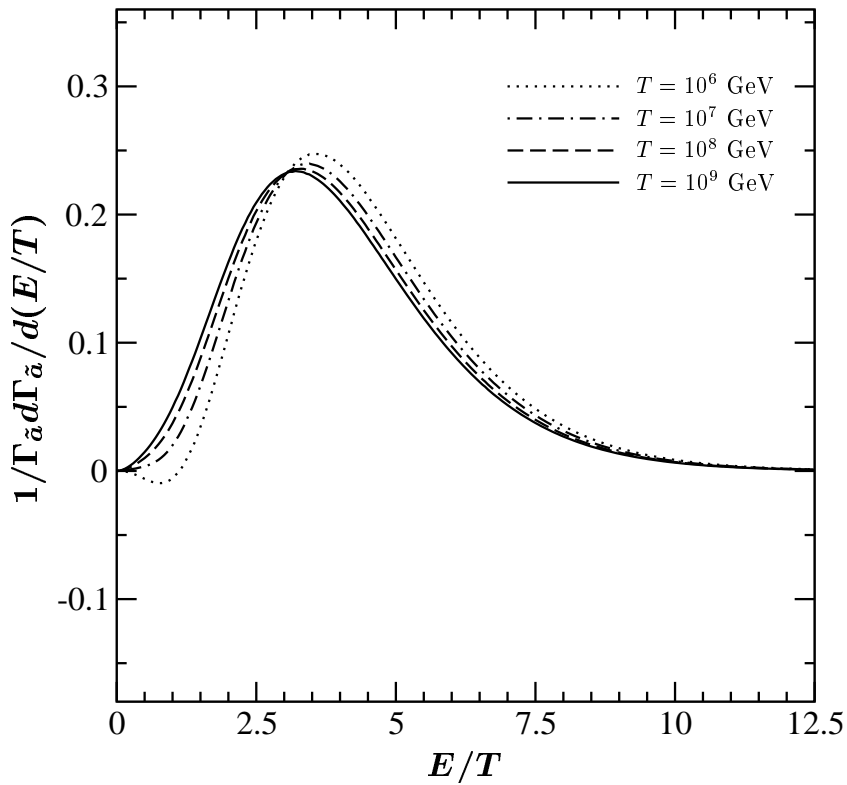

Figure 3: The normalized thermal axino production rate $1/\Gamma_{\tilde{a}}\, d\Gamma_{\tilde{a}}/d(E/T)$ as a function of $E/T$ for $N_c = 3$, $n_f = 6$, and temperatures of $T = 10^6$ GeV (dotted line), $T = 10^7$ GeV (dash-dotted line), $T = 10^8$ GeV (dashed line), and $T = 10^9$ GeV (solid line). The results are obtained from (3.30) derived for axino energies $E \gtrsim T$.

# Acknowledgements

We thank Josef Pradler for noticing the error in the calculation of thermal gravitino production, which was adopted in our calculation of thermal axino production.